\documentclass[%
    aps,%
    prl,%
    twocolumn,%
    showpacs,%
    footinbib,%
    superscriptaddress%
]{revtex4-1}

\usepackage{graphicx}

\usepackage{dcolumn}
\usepackage{color}

\usepackage{bm}

\usepackage{physics}

\usepackage{xr-hyper}

\usepackage{hyperref}
\hypersetup{%
	colorlinks=true,%
	linktocpage=true,%
	breaklinks=true,%
	pageanchor=true,%
	bookmarksnumbered,%
	bookmarksopen=true,%
	bookmarksopenlevel=1,%
	hypertexnames=true%
}

\usepackage[capitalise]{cleveref}

\def\k{\mathbf{k}}
\newcommand\avg[1]{\left<#1\right>}

\begin{document}

\title{Supersolid stripe crystal from finite-range interactions on a lattice}

\author{Guido Masella}
\affiliation{%
    icFRC, IPCMS (UMR 7504) and ISIS (UMR 7006), Universit\'e de Strasbourg and
    CNRS, 67000 Strasbourg, France
}

\author{Adriano Angelone}
\affiliation{%
    Abdus Salam International Centre for Theoretical Physics, strada Costiera
    11, 34151 Trieste, Italy
}
\affiliation{%
    SISSA, via Bonomea 265, 34136 Trieste, Italy
}

\author{Fabio Mezzacapo}
\affiliation{%
    Univ Lyon, Ens de Lyon, Univ Claude Bernard, CNRS, Laboratoire de
    Physique, F-69342 Lyon, France
}
\affiliation{%
    icFRC, IPCMS (UMR 7504) and ISIS (UMR 7006), Universit\'e de Strasbourg and
    CNRS, 67000 Strasbourg, France
}

\author{Guido Pupillo}
\affiliation{%
    icFRC, IPCMS (UMR 7504) and ISIS (UMR 7006), Universit\'e de Strasbourg and
    CNRS, 67000 Strasbourg, France
}

\author{Nikolay V. Prokof'ev}
\affiliation{%
    Department of Physics, University of Massachusetts, Amherst, Massachusetts
    01003, USA
}
\affiliation{%
    icFRC, IPCMS (UMR 7504) and ISIS (UMR 7006), Universit\'e de Strasbourg and
    CNRS, 67000 Strasbourg, France
}

\date{\today}

\begin{abstract}
    Strong, long-range interactions present a unique challenge for the
    theoretical investigation of quantum many-body lattice models, due to the
    generation of large numbers of competing states at low energy. Here, we
    investigate a class of extended bosonic Hubbard models with off-site terms
    interpolating between short- and infinite-range,  thus allowing for an
    exact numerical solution for all interaction strengths. We predict a novel
    type of stripe crystal at strong coupling. Most interestingly, for
    intermediate interaction strengths we demonstrate that the stripes can
    turn superfluid, thus leading to a self-assembled array of quasi
    one-dimensional superfluids. These bosonic \textit{superstripes} turn into
    an isotropic supersolid with decreasing the interaction strength. The
    mechanism for stripe formation is based on cluster self-assemblying in the
    corresponding classical ground state, reminiscent of classical soft-matter
    models of polymers, different from recently proposed mechanisms for cold
    gases of alkali or dipolar magnetic atoms.
\end{abstract}


\maketitle
The effects of long-range interactions on quantum phases of many-body lattice
systems is a hot topic of research~\cite{Campa2014, Kastner2010, Kastner2011,
Lahaye2009, Baranov2012}, which is driven by outstanding advances in precision
experiments with strongly interacting magnetic atoms~\cite{Schmitt2016,
Chomaz2018, Vaidya2018, Lepoutre2018, Lucioni2018}, polar
molecules~\cite{Carr2009, Jin2012}, Rydberg-excited atoms~\cite{Saffman2010,
Low2012}, ions~\cite{Britton2012, Bermudez2013, Richerme2014, Jurcevic2014},
and neutral atoms coupled to photonic modes~\cite{John1990, Schneider2012,
Shahmoon2013, Ritsch2013, Tang2018a, Douglas2015}.  For bosonic particles,
exact numerical results from quantum Monte-Carlo methods can in principle
predict thermodynamic properties of any unfrustrated model.  However,
long-range interactions in combination with confinement to periodic potentials
present a unique challenge as they generate a proliferation of metastable
low-energy states, whose number exponentially increases with the system
size~\cite{Menotti2007,Trefzger2008}, even in the absence of external
frustration.  This usually results in, e.g., devil's staircase-type structures
that are essentially intractable~\cite{Bak1982, Burnell2009,
Capogrosso-Sansone2010, Pollet2010}.

Much interest was recently generated by the demonstration of stripe behavior
in spin-orbit-coupled Bose-Einstein condensates~\cite{Li2013, Cong-Jun2011,
Ho2011, Wang2010, Li2017} as well as droplet formation in clouds of dipolar
magnetic atoms in the mean-field regime~\cite{Kadau2016, Ferrier-Barbut2016,
Chomaz2016}, due to a competition of quantum fluctuations, short- and
long-range interactions~\cite{Ferrier-Barbut2016, Wachtler2016, Baillie2016}.
Exact numerical results have further demonstrated theoretically that
anisotropic dipolar interactions for particles confined to two dimensions (2D)
can generate stripe behavior while preserving superfluidity~\cite{Macia2012,
Bombin2017},  corresponding to a possible realization of so-called stripe
supersolidity~\cite{Batrouni2000, Boninsegni2012}.  The latter has a long
history in quantum condensed matter, where it was first introduced as
\textit{superstripe} phase, in that non-homogeneous metallic structures with
broken spatial symmetry were found to favor
superconductivity~\cite{Bianconi2013}.
While the microscopic origin of such a phase is still a subject of debate, it
is clear that a key role is played by a combination of strong interactions and
the lattice potential. In this context, key open challenges are to propose and
understand basic mechanisms for (super)stripe formation  on lattice geometries
of experimental interest and to provide exact theoretical predictions in the
regime of strong interactions.

\begin{figure}[t]
  \includegraphics[width=1.0\columnwidth]{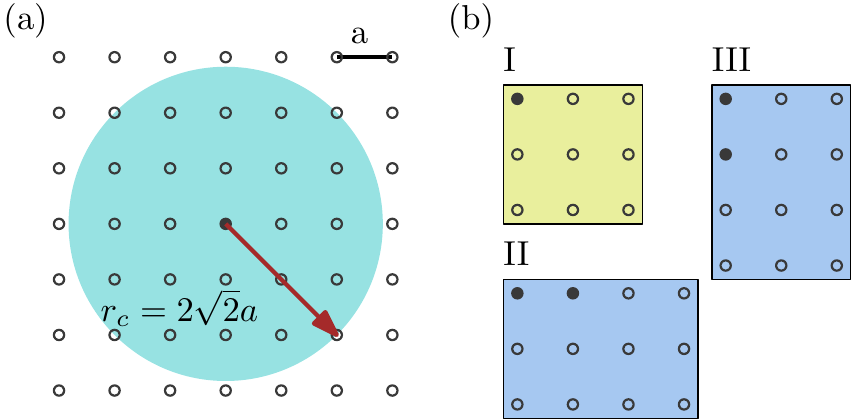}
    \caption{\label{fig:model}
        Sketch of the interaction potential chosen in our work on a square
        lattice of spacing $a$ [panel~(a)]. The shaded region indicates the
        interaction range, which extends up to the critical radius $r_c =
        2\sqrt{2}a$. In the large-$V$ limit, the GS of model
        \cref{eq:hamiltonian} at our chosen density $\rho=5/36$ can be
        found by tiling the lattice with clusters of type I, II and III
        [panel~(b)].  Empty and full circles refer to empty and occupied
        lattice sites, respectively.
    }
\end{figure}
\begin{figure*}[t]
  \includegraphics[width=2.0\columnwidth]{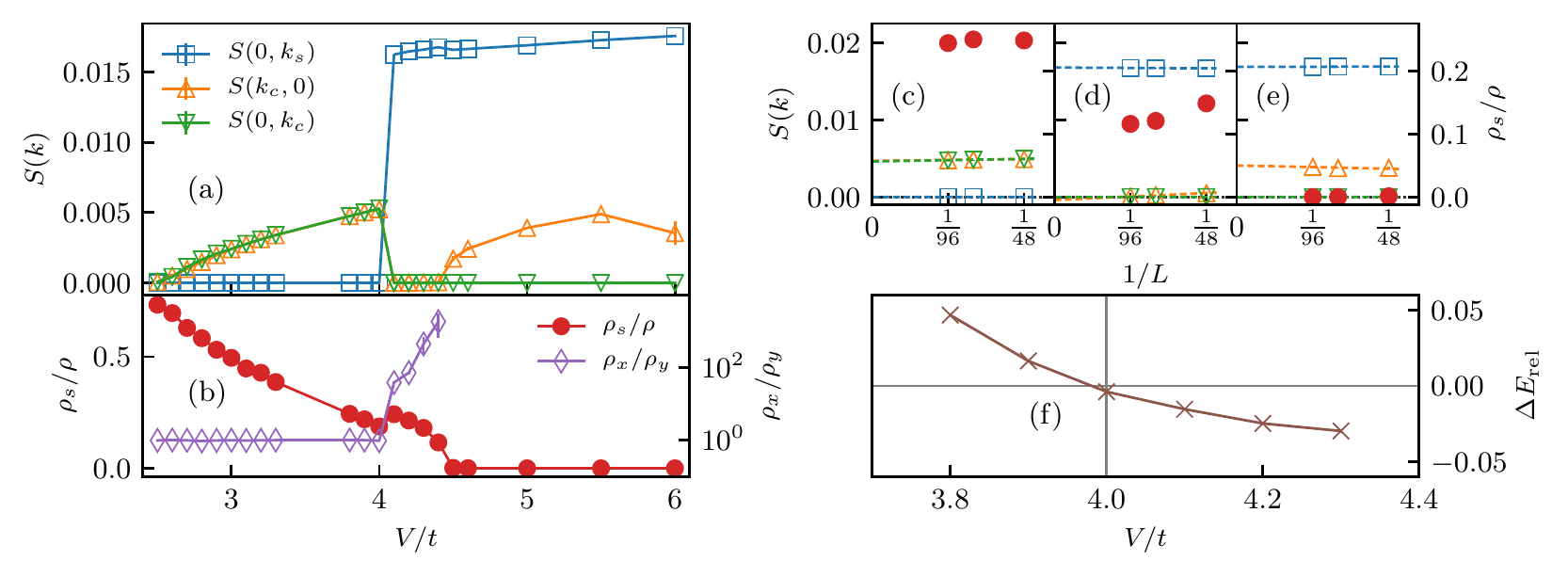}
    \caption{\label{fig:orderparams}
        Panel~(a): Structure factor $S(\k)$ as a function of $V/t$ for values
        of lattice wave vector $\k= (k_c, 0)$ (up-pointing triangles),
        $(0, k_c)$ (down-pointing triangles), $(0, k_s)$ (squares)
        characteristic of the IS, SC, and SS ordered phase, respectively (see
        text). Here $k_c = 2\pi \times 7/24$ and $k_s = 2\pi\times 1/3$.
        Panel~(b): Superfluid fraction $\rho_s/\rho$ (circles), and ratio
        between the superfluid responses $\rho_x/\rho_y$ along the $x$
        horizontal and $y$ vertical axis (diamonds). Panels~(c)-(e):
        Finite-size scaling of $S(\k)$ and $\rho_s/\rho$ for $V/t=3.7$
        [panel~(c)], $4.4$ [panel~(d)] and $6.0$ [panel~(e)]. Here same
        symbols refer to the same observables of panels (a) and (b) while
        lines are linear fitting functions to our numerical data. Panel~(f):
        Relative energy difference $\Delta E_{\rm rel}$ between the SS and IS
        phases as a function of $V/t$. The SS has lower energy where $\Delta
        E_{\rm rel}$ is negative. In panels~(a), (b), and (f) $L=96$, in all
        panels $T/t=1/20$.
    }
\end{figure*}

In this work,  we study the low temperature phase diagram of an ensemble of
bosonic particles confined to a square lattice geometry and interacting via a
soft-shoulder potential. The latter has a finite range that includes several
lattice sites, thus interpolating between nearest-neighbor and long-range
physics, while remaining numerically tractable. We utilize numerically exact
quantum Monte-Carlo simulations to study the phase diagram of this system as a
function of the interaction strength, finding several novel features: (i) For
sufficiently strong interactions, the ground state (GS) is a highly anisotropic,
insulating stripe crystal that emerges due to cluster self-assembling in the
corresponding classical GS. For intermediate interaction strength,
we find a surprising (ii) supersolid-supersolid quantum phase transition that
separates an isotropic supersolid state from (iii) a highly anisotropic stripe
state. In the latter, superfluidity mostly occurs along horizontal (vertical)
stripes and is not suppressed at the supersolid-supersolid transition, while
diagonal long-range order is found in the perpendicular direction --
reminiscent of the so-called \textit{superstripe} phase found in lattice-based
superconductors~\cite{Bianconi2013}.

We consider the following extended Hubbard Hamiltonian for hard-core bosons
confined to 2D
\begin{equation}\label{eq:hamiltonian}
    \mathcal{H} = -t \sum_{\left<i,j\right>} \left(b^{\dagger}_i b_j +
    \text{h.c.}\right) +V \!\!\!\!\sum_{i<j ; r_{ij} \le r_c}\!\!\! n_i n_j,
\end{equation}
where $b_j$ ($b^\dagger_j$) is the bosonic annihilation (creation) operator on
site $j$, $n_j = b^\dagger_j b_j$, and $t$ is the nearest-neighbor hopping
energy on a square lattice of spacing $a$. In the following $t$ and $a$ are
taken as units of energy and length, respectively. The last term of
\cref{eq:hamiltonian} represents the soft-shoulder interaction between
bosons with strength $V$, $r_{ij}$ is the distance between sites $i$ and $j$,
with $r_c$ the interaction potential cutoff.
In classical physics, this interaction is of interest for soft-matter models
of, e.g., colloids~\cite{Mladek2006a, Lenz2012, Sciortino2013}. In quantum
physics, similar interactions can be engineered in clouds of cold Rydberg
atoms, by weakly-admixing a Rydberg state to the GS using laser
light~\cite{Henkel2010, Pupillo2010, Johnson2010, Cinti2010, Honer2010,
Jau2016, Zeiher2016}.
Here, we choose $r_c = 2 \sqrt{2}a$ for which in the classical limit $V/t
\rightarrow \infty$ each particle tries to establish an avoided square region
of total area $16a^2$ [see \cref{fig:model}(a)]. For density $\rho = 1/9$
this is indeed possible and the system can arrange into an optimal
configuration characterized by zero potential energy by covering the lattice
with clusters of type I, see \cref{fig:model}(a). Conversely, for higher
densities the classical GS corresponds to the solution of a tiling
problem, where tiles are constituted by clusters of particles and holes that
are effectively bound together by the repulsive interactions. The number of
such clusters, or tiles, increases with increasing particle density.
Figure~\ref{fig:model}(b) shows the three clusters I, II, and III (i.e., the
tiles) that appear at low energy for densities $1/9 < \rho < 1/6$. Similarly
to the 1D case~\cite{Mattioli2013, Dalmonte2015}, the classical GS
can then be built from  the exponentially large number of permutations of
clusters I-III. This large degeneracy is characteristic of long-range
interactions and can in principle constitute an obstacle to the solution of
the quantum problem. In the following, we determine the effects of quantum
fluctuations on this highly degenerate classical GS by computing the
quantum phase diagram for model~\cref{eq:hamiltonian}, for an example
density $\rho = 5/36$. Our focus is the determination of quantum phases and
phase transitions in this system.

We study Hamiltonian \cref{eq:hamiltonian} by means of path integral
Quantum Monte Carlo simulations based on the worm
algorithm~\cite{Prokofev1998}. This technique is numerically exact for
unfrustrated bosonic models, giving access to accurate estimates of
fundamental observables such as, e.g., the superfluid fraction
$\rho_s/\rho=\avg{W_x^2 + W_y^2}/\left(4t\beta\rho\right)$, the static
structure factor $S(\mathbf{k})=\avg{\sum_{ij} e^{-i\mathbf{k} \cdot
\mathbf{r}_{ij}}n_i n_j}/N^2$, and the single-particle equal time Green's
function $G(\mathbf{r}) = \left<\sum_i b^\dagger_ib_{i+\mathbf{r}}\right>/N$.
They measure superfluidity, diagonal long-range, and off-diagonal long-range
orders, respectively. Here, $\beta =1/(k_B T)$ is the inverse temperature,
with $k_B$ the Boltzmann constant (set to unity); $W_x$ ($W_y$) is the winding
number along the $x$ ($y$) direction, $\mathbf{k}$ is a lattice wave vector
and $\avg{\dots}$ stands for statistical averaging. Calculations are performed
on lattices of size $N=L \times L$, with $L$ as large as $L=96$ and
temperatures as low as $T/t= 1/96$. We find that for $T/t \le 1/20$ results
are essentially indistinguishable from the extrapolated GS ones.
\begin{figure}[t]
  \includegraphics[width=1.0\columnwidth]{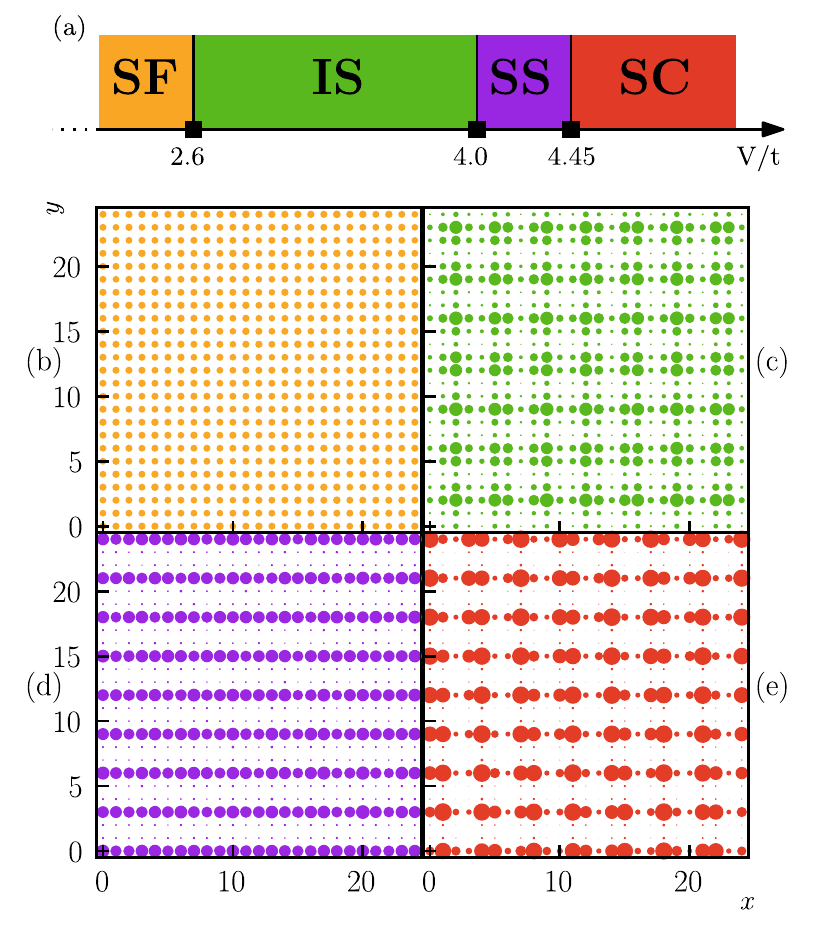}
    \caption{\label{fig:phases}
        Ground state (GS) phase diagram of model \eqref{eq:hamiltonian} as a
        function of the interaction strength $V/t$ [panel (a)]. For increasing
        values of $V/t$ the GS is a superfluid (SF), an isotropic
        supersolid (IS), an anisotropic stripe supersolid (SS) and stripe
        crystal (SC) (see text).  Panels (b), (c), (d), (e) show site-density
        maps of a portion of the system for representative interaction
        strengths at which the GS is a SF ($V/t=2.5$), IS
        ($V/t=3.7$), SS ($V/t=4.3$), and SC ($V/t=6.0$), respectively.
        The size of the dots is proportional to the occupation of the
        corresponding sites.
    }
\end{figure}

The main results are presented in \cref{fig:orderparams}. Panels~(a) and
(b) show estimates of the structure factor $S(\k)$ and the superfluid fraction
$\rho_s/\rho$ (left ordinate axis) together with the ratio between the
superfluid responses $\rho_x/\rho_y \equiv \avg{W_x^2}/\avg{W_y^2}$ along the
horizontal and vertical directions (right ordinate axis) as a function of
$V/t$, for $T/t=1/20$ and $N=96 \times 96$, respectively. Examples of
finite-size scalings [panels~(c)-(e) of the same figure] clarify that the
chosen system size is large enough to provide an accurate description of the
various observables in the thermodynamic limit, as results obtained for $L=96$
essentially coincide with the extrapolated estimates. The combination of
$S(\k)$, $\rho_s/\rho$, and their anisotropies allows for the determination of
the quantum phases.

We find that for weak interaction strengths $V/t \lesssim 2.6$ the GS
is a homogeneus superfluid (SF) with $\rho_s/\rho > 0$,
$\rho_x\simeq\rho_y$, and $S(\k)=0$.  For $2.6 \lesssim V/t \lesssim 4.45$,
however, both the superfluid fraction and the structure factor are finite,
indicating the presence of a supersolid GS.  Surprisingly, in this
range of interaction strength we find two distinct supersolids. Specifically,
an isotropic supersolid (IS) and an anisotropic stripe supersolid (SS) occur
for $2.6 \lesssim V/t \lesssim 4.0$ and $4.0 \lesssim V/t \lesssim 4.45$,
respectively. Within the IS phase, $S(\k)$ [\cref{fig:orderparams}(a)]
takes its maximum value for $\mathbf{k} = (0,\pm k_c)$ and $(\pm k_c,0)$
(down- and up-pointing triangles, respectively) with $k_c= 2\pi \times 7/24$;
in IS the superfluid response is isotropic [\cref{fig:orderparams}(b),
diamonds].
\begin{figure}[t]
  \includegraphics[width=1.0\columnwidth]{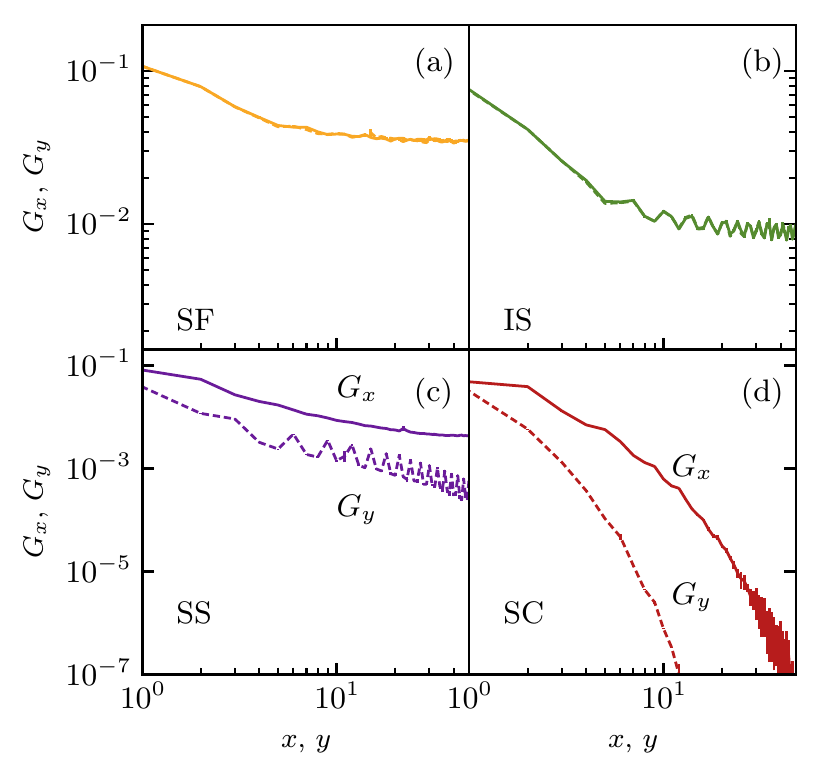}
    \caption{\label{fig:greenfunctions}
        Green's functions $G_x$ and $G_y$ along the $x$ (solid lines) and $y$
        (dashed lines) directions, respectively, in the SF, IS, SS, and SC\@.
        Here $L=96$, $T/t=1/20$ and the values of $V/t$ are the same as those
        in \cref{fig:phases}(b)-(e).
    }
\end{figure}
In contrast, in the SS phase the diagonal long-range order is found along one
direction only (i.e., the $y$ direction in the figure),  being drastically
suppressed along the perpendicular one; in addition, the maximum of the
structure factor is found for $\mathbf{k} = (0,\pm k_s)$, with $k_s =
2\pi\times 1/3 \neq k_c$, while $S(0,\pm k_c) = S(\pm k_c,0) =0$
[\cref{fig:orderparams}(a), triangles]. Here, the superfluid response
becomes strongly anisotropic with $\rho_x \gg \rho_y$, signalling the
formation of superfluid stripes along the $x$-axis. This corresponds to a
transition to a self-assembled array of essentially one-dimensional
superfluids, which, unexpectedly, have larger superfluid density near the
phase boundary: $\rho_s/\rho$ initially increases within the SS phase, before
decreasing again, with increasing $V/t$.

Finally, for $V/t \gtrsim 4.45$ the system loses its superfluid character and,
although the maximum value of $S(\k)$ still occurs for $\k=(0, \pm k_s)$,
secondary peaks emerge at $\k=(\pm k_c, 0)$. These latter peaks imply both
crystallization along the stripe direction as well as the emergence of weak
correlations between particles across different stripes (see below). The
resulting GS is a normal crystal.

We find that different metastable states with entirely different quantum
orders compete in the region of intermediate strengths of interactions $V/t
\gtrsim 3.0$. In order to determine the GS we perform two sets of
simulations: namely, starting from the equilibrium configuration at
$V_1/t=3.7$ ($V_2/t=5.0$) a careful annealing in the interaction strength is
performed with annealing step $0< \Delta V_1/t \leq 0.1$ ($-0.1 \leq \Delta
V_2/t <0$).  When the desired target value of $V/t$ is reached the calculation
leading to a lower energy $E$ is taken as the GS.
Figure~\ref{fig:orderparams}(f) shows the relative energy difference $\Delta
E_{\mathrm rel} = (E_{2} - E_{1})/E_{1}$ as a function of $V/t$. The change in
sign at $V/t\simeq4.0$ signals the phase transition between the IS and SS
phases.  The corresponding sudden changes in crystalline order, measured by
discontinuities of the structure factors in \cref{fig:orderparams}(a), are
consistent with a first order phase transition between the two supersolids.
\begin{figure}[t]
  \includegraphics[width=0.6\columnwidth]{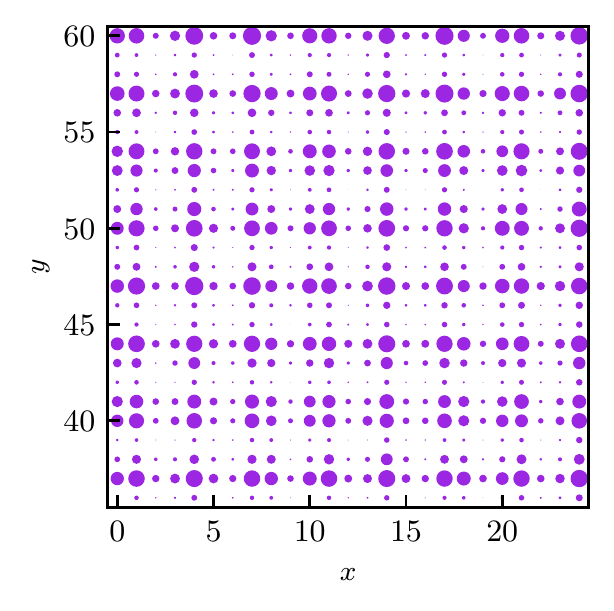}
    \caption{\label{fig:quenches}
        Out-of-equilibrium density snapshot for a system with $L=96$, $V/t =
        4.1$ after a quench down to target temperature $T/t=1/20$. For this
        choice of parameters the corresponding equilibrium phase is a SS. The
        size of the dots is proportional to the averaged occupation of the
        corresponding sites.
    }
\end{figure}

The formation of stripes remains favored for larger $V/t$: hence, the phase
transition from the SS to the stripe crystal phase at $V/t \simeq 4.45$ is
resolved by monitoring the vanishing of superfluidity fraction. \\

The GS phase diagram of model~\eqref{eq:hamiltonian} is
summarized in \cref{fig:phases}. The demonstration of the existence of
novel superfluid and insulating stripe crystals, as well as an exotic
supersolid-supersolid quantum phase transitions due to classical cluster
formation in a rather general model with a simple isotropic interaction are
the main results of this work. Further insight into the discussed ground
states can be obtained from the density maps in \cref{fig:phases}(b)-(e)
and the corresponding Green's function $G_x$ ($G_y$) along the $x$ ($y$) axis
[\cref{fig:greenfunctions}]. As expected, for small $V/t$ (where the
system is a homogeneous SF) the average occupation number at each site equals
the density $\rho$ [\cref{fig:phases}(b)].  Similarly, the Green's
functions (GF) are equal in the $x$ and $y$ directions at all distances (i.e.,
$G_x\simeq G_y$), within the statistical error bars. They become nearly flat
at large distances, which is consistent with the presence of off-diagonal
(quasi) long-range order [\cref{fig:greenfunctions}(a)]. In the IS phase
[\cref{fig:phases}(c)] the isotropic ordered structure formed by clusters
of particles coexists with quantum exchanges and superfluidity. Here,
$G_{x,y}$ displays a weak power-law decay, indicating off-diagonal
quasi-long-range order, accompanied by oscillations
[\cref{fig:greenfunctions}(b)], which we find to have a periodicity
consistent with particle exchanges between different clusters, thus
identifying the underlying classical structure. When stripes are formed in the
SS phase, \cref{fig:phases}(d), no density modulations appear along their
direction (i.e., the horizontal one). $G_x$ is found to decay as a power-law
[\cref{fig:greenfunctions}(c)], consistent with the measured finite
superfluidity along the stripe direction. Long exchange cycles of identical
particles take place almost exclusively along the stripes, being strongly
suppressed in the perpendicular direction. The overall picture here is that of
a 2D quantum system of quasi 1D superfluids (i.e. the stripes). In the SC
phase the emergence of a nearly classical cluster-crystalline structure is
evident [\cref{fig:phases}(e)]. Here long quantum exchanges are completely
suppressed and clusters (and particles) can only slightly fluctuate around
their equilibrium position due to zero point motion, implying an exponential
decay of the GF's in \cref{fig:greenfunctions}(d), albeit with different
slopes in the $x$ and $y$ directions.\\

We find that the phases above are robust for density variations within the
range $1/9<\rho<1/6$, where clusters I-III appear at low energy for large
$V/t$.  In the Supplementary Material~\cite{Note1} we demonstrate that they
also persist for $\rho=1/6$, where only clusters of type II and III exist for
stong interactions, albeit with a different cluster periodicity (i.e.
different $k_c$).\\

Finally, to exemplify possible out-of-equilibrium scenarios that can emerge
with imperfect annealing,  \cref{fig:quenches} shows a density snapshot
obtained when the system is driven away from equilibrium via a temperature
quench. Here the target temperature is $T/t=1/20$ for a value of $V/t$ at
which the equilibrium phase is a SS. The resulting snapshot is isotropic
rather than anisotropic with a crystalline structure similar to panel~(c) in
\cref{fig:phases}, where diagonal long-range order is found for
characteristic wave vectors $\k = (0,\pm k^*)$ and $(\pm k^*,0)$ with $k^*
\neq k_c, k_s$, and the value of $\rho_s/\rho$ is much smaller than the
equilibrium one.

Increasing $r_c$ or smoothening the edges of the interactions can effectively
result in the inclusion of more sites in the interaction volume~\cite{Note1}.
This can change the number and type of clusters that appear at low-energy, and
thus the resulting crystal structures. For example, for a larger $r_c=3a$ and
$\rho=1/7$ the strong-coupling phase is a SC not oriented in the $x$ or
$y$-directions. In these cases, we find that annealing can become increasingly
difficult as equilibration is often dominated by the presence of many
metastable states, typical of long-range models.  A detailed investigation of
metastability for model~\eqref{eq:hamiltonian} is presented
in~\cite{Angelone2016a}.

We have demonstrated that stripe supersolid and crystals may be realized in
the ground state of bosonic, frustration-free cluster-forming Hamiltonians. In
particular, for intermediate interaction strength the competition between
quantum fluctuations and cluster formation gives rise to a novel
supersolid-supersolid transition between an isotropic cluster supersolid and
an anisotropic stripe supersolid. These results exemplify the complexities of
the determination of quantum phase diagrams of systems with long-range
interactions, in a regime where calculations are still feasible.  Intriguing
out-of-equilibrium scenarios may also emerge and will be the subject of future
investigations~\cite{Angelone2016a}. Our predictions should be of direct
interest for experiments with cold Rydberg-dressed atoms in an optical
lattice~\cite{Schempp2015, Boulier2017}. More generally, they constitute a
step towards the understanding of how long-range interactions can affect the
properties of ultracold gases.\\

\footnotetext[1]{%
    See Supplemental Material for discussion on the effects of variations in
    density and interaction range on the many-body phases presented here,
    which includes Refs.~\cite{Cinti2010, Zeiher2016, Glaetzle2015,
    vanBijnen2015}
}

\begin{acknowledgments}
    \emph{Acknowledgments} --
    We are grateful to Marcello Dalmonte, Tao Ying, and Peter Zoller for many
    insightful discussions. Work in Strasbourg was supported by the ERC
    St-Grant ColDSIM (No. 307688), ANR-“ERA-NET QuantERA” - Projet “RouTe”
    (ANR-18-QUAN-0005-01), and additional ANR-FWF grant BLUESHIELD. A. A.
    acknowledges partial support by the ERC under Grant No. 758329 (AGEnTh).
    F. M. acknowledges support from F\'ed\'eration de Recherche Andr\'e Marie
    Amp\`ere (FRAMA). N. P.  acknowledges support from the MURI Program
    “Advanced quantum materials—a new frontier for ultracold atoms” from AFOSR
    and the National Science Foundation under the Grant No. DMR-1720465.
    Computing time provided by HPC-UdS.

    \emph{Note added} --
    After the acceptance of this work, we became aware of a recent
    study~\cite{Morita2019} showing transitions between supersolid phases in a
    different model.  However, as in previous cases (see
    e.g.~\cite{Bruder1993}), the mechanism behind the transition is a change
    in density, rather than cluster formation induced by the interaction
    potential at fixed density.
\end{acknowledgments}

\bibliography{library.bib}

\end{document}


\title{Supplementary material of\\ ``Supersolid stripe crystal from finite-range
interactions on a lattice''}

\author{Guido Masella}
\affiliation{%
    icFRC, IPCMS (UMR 7504) and ISIS (UMR 7006), Universit\'e de Strasbourg and
    CNRS, 67000 Strasbourg, France
}

\author{Adriano Angelone}
\affiliation{%
    Abdus Salam International Centre for Theoretical Physics, strada Costiera
    11, 34151 Trieste, Italy
}
\affiliation{%
    SISSA, via Bonomea 265, 34136 Trieste, Italy
}

\author{Fabio Mezzacapo}
\affiliation{%
    Univ Lyon, Ens de Lyon, Univ Claude Bernard, CNRS, Laboratoire de
    Physique, F-69342 Lyon, France
}
\affiliation{%
    icFRC, IPCMS (UMR 7504) and ISIS (UMR 7006), Universit\'e de Strasbourg and
    CNRS, 67000 Strasbourg, France
}

\author{Guido Pupillo}
\affiliation{%
    icFRC, IPCMS (UMR 7504) and ISIS (UMR 7006), Universit\'e de Strasbourg and
    CNRS, 67000 Strasbourg, France
}

\author{Nikolay V. Prokof'ev}
\affiliation{%
    Department of Physics, University of Massachusetts, Amherst, Massachusetts
    01003, USA
}
\affiliation{%
    icFRC, IPCMS (UMR 7504) and ISIS (UMR 7006), Universit\'e de Strasbourg and
    CNRS, 67000 Strasbourg, France
}

\date{\today}

\begin{abstract}
    We discuss the effects of variations in density and interaction range on
    the many-body phases of~\paper{}. In particular, for a choice of density
    $\rho=1/6$, we demonstrate the existence of a stripe crystal for large
    values of interaction strength $V/t$ and of a quantum phase transition
    from an anisotropic supersolid to a homogeneous supersolid with
    decreasing $V/t$ for Hamiltonian \paperhamiltonian{} in the main
    text \paper{}.
\end{abstract}

\maketitle

The numerical results presented in \paper{}~are obtained for a specific choice
of density $\rho=5/36$ and for a soft-shoulder interaction potential with
range $r_c=2\sqrt{2}$. Below we show that the main quantum phases and phase
transitions found in \paper{}~are robust for density variations within the
range $1/9<\rho\leq1/6$, where clusters of type II and III (in addition to
type I) fully determine the crystalline structure for large interaction
strengths $V/t$, consistent with the discussion in the main text.  This is
explicitly shown below for the limiting case $\rho = 1/6$.

Variations of Hamiltonian terms such as the interaction range $r_c$ or shape
can in principle lead to different crystalline phases at strong coupling with
respect to the stripe crystal found in the main text. Below we discuss the
crystalline phases occurring by increasing $r_c$ or by considering a smooth
interaction potential instead of the step-like potential of the main text.

\section{Quantum phases of Hamiltonian (1) for density $\rho=1/6$}

The phases  of \paperhamiltonian{} described in the main text for
$\rho=5/36$ are representative of densities in the range $1/9<\rho<1/6$, where
the clusters of type I, II and III in~\papermodel{} dominate the
dynamics for a large ratio $V/t$. Here, we investigate the case $\rho = 1/6$,
corresponding to the smallest density increase for which the classical tiling
problem has a different structure, as only the clusters of type II and III
occur for large $V/t$.  Calculations are performed for system sizes up to
$N=96\times96$ and temperatures as low as $T/t=1/20$.\\

\begin{figure}[tb]
    \centering
    \includegraphics[width=1.0\columnwidth]{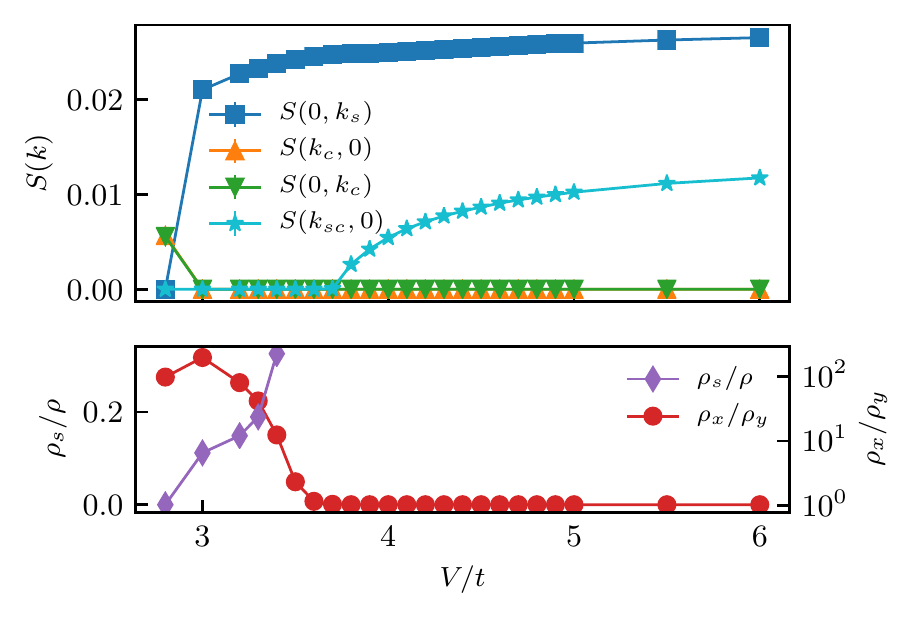}
    \caption{\label{fig:orderpars}%
        Panel (a): Structure factor $S(\k)$ as a function of $V/t$ for values
        of lattice wave vector $\k= (k_c, 0)$ (up-pointing triangles), $(0,
        k_c)$ (down-pointing triangles), $(0, k_s)$ (squares), $(k_{sc}, 0)$
        (stars) characteristic of the IS, SS, and SC ordered phase,
        respectively (see text). Here $k_c = 2\pi \times 7/24$ and $k_s =
        2\pi\times 1/3$. Panel (b): Superfluid fraction $\rho_s/\rho$
        (circles), and ratio between the superfluid responses $\rho_x/\rho_y$
        along the $x$ horizontal and $y$ vertical axis (diamonds).
        Data taken from calculations with $N=96\times96$, $\rho=1/6$, and
        $T/t=1/20$.
    }
\end{figure}

The main results are presented in \cref{fig:orderpars}, where, following the
definitions found in \paper{}, the structure factor $S(\k)$
[panel (a)], the superfluid fraction $\rho_s/\rho$ [panel (b) left ordinate
axis], and the ratio between the superfluid responses $\rho_x/\rho_y$ [panel
(b) right ordinate axis] are plotted as a function of $V/t$, for $T/t=1/20$ and
$N=96\times96$.

For large enough interaction strengths $V/t > 3.9$, we find a stripe crystal
(SC) with no superfluidity, similar to the case treated in the main text.
However, the peaks in $S(\k)$ occur here for $\k = (0,\pm k_s)$ and $\k =
(k_{sc}, 0)$, with $k_s = 2\pi/3$ and  $k_{sc} = 2\pi/4$, respectively. The
crystalline order appearing on the $x$-axis is thus slightly different from
that of the SC phase in \paper{} ($k_{sc} \neq k_c = 2\pi\times 7/24$).

\begin{figure}[htb]
    \centering
    \includegraphics[width=1.0\columnwidth]{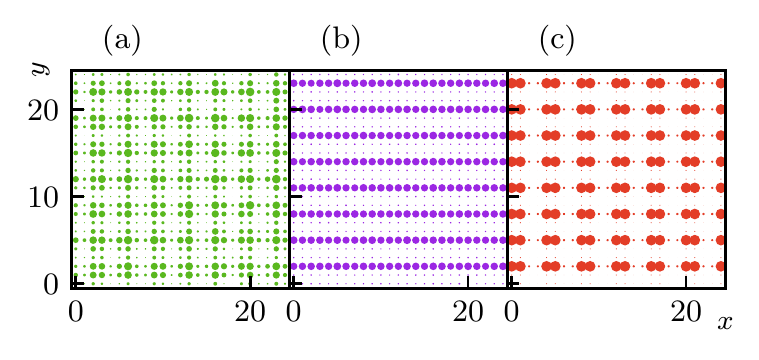}
    \caption{\label{fig:phases}%
        Site-density maps of a portion of the system for representative
        interaction strengths at witch the ground state is a IS ($V/t=2.8$)
        [panel (a)], SS ($V/t=3.2$) [panel (b)], and SC ($V/t=6.0$) [panel
        (c)]. The size of the dots is proportional to the occupation number of
        the corresponding lattice site. The data is taken from calculations
        at $T/t = 1/20$, $\rho=1/6$, and $N=96\times96$.
    }
\end{figure}

Similarly to the main text, by decreasing the interaction strengths we first
find an anisotropic stripe supersolid (SS) and then an isotropic supersolid
(IS) phase.  The latter phases occur for interaction strengths  $3.0 \lesssim
V/t \lesssim 3.9$ and $V/t \lesssim 3$, respectively.  The peaks in the
structure factor within the IS phase are at $\k = (\pm k_c, 0)$ and $\k = (0,
\pm k_c)$, identical to the case $\rho=5/36$ discussed in the main text. \\

Further insight can be obtained from the density maps in \cref{fig:phases}. In
particular, panel (c) shows that for large $V/t$ in the classical regime the
ground state corresponds to a perfect crystalline tiling of the surface with
just one single kind of clusters (i.e., cluster II in the $x-$direction in the
figure). This results in a SC that is qualitatively similar to the one of
\paper{}, albeit with a slightly different periodicity in the $x$-direction, as
discussed above.  By decreasing $V/t$, a melting of this crystalline phase
gives way to the SS and then IS phases.\\

In conclusion, we find that the main results discussed in \paper{}, namely the
appearance of an anisotropic (stripe) supersolid and a transition between two
different supersolids (IS and SS), are robust against different choices of
density in the range $1/9<\rho\leq 1/6$, as long as the strong-coupling
low-energy physics is dominated by clusters of the type II and III
in~\papermodel{}, as expected. Further increasing densities with
$\rho >1/6$ introduces different kinds of clusters in the classical ground
state that alter both the strong-coupling stripe crystalline structure, and
the ensuing supersolid phases for intermediate values of $V/t$. The
corresponding quantum phases have then to be investigated on a case-by-case
basis, for each density.

\section{Effects of the interaction range and shape on the crystal phases of Eq.~(1)}

\begin{figure}[htbp]
    \centering
    \begin{minipage}[t]{0.4\columnwidth}
        \vspace{0pt}
        (a)\\
        \includegraphics[width=\linewidth]{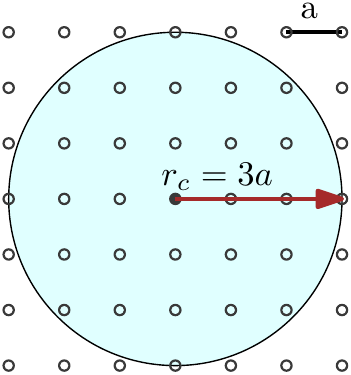}
    \end{minipage}
    \hfill
    \begin{minipage}[t]{0.55\columnwidth}
        \vspace{0pt}
        (b)\\
        \includegraphics[width=\linewidth]{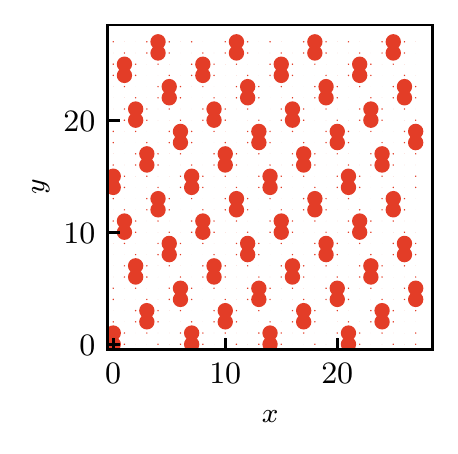}
    \end{minipage}
    \caption{\label{fig:rc3}%
        Panel (a): Interaction volume corresponding to the model
        \paperhamiltonian{} with the cutoff radius of the interaction
        being $r_c=3.0$.  Panel (b): Side-density map of a system with
        $N=28\times28$ showing a possible anisotropic crystalline structure
        forming for the choice $r_c = 3.0$.  The size of the dots is
        proportional to the occupation number of the corresponding lattice
        site. The data is taken from calculations at $T/t = 1/20$ and $V/t =
        6.0$ at the fixed density $\rho = 1/7$.
    }
\end{figure}

Varying the interaction range $r_c$ in \paperhamiltonian{} can change
the number and shape of relevant clusters in the large $V/t$ limit. As a
consequence, the many-body ground state can change from the predictions of
\paper{}. As an example, \cref{fig:rc3}(b) shows a site-density map of a normal
(not superfluid) crystalline phase appearing at $V/t = 6$ for a choice of
interaction range $r_c = 3.0$ and density $\rho = 1/7$. The stripe crystal
structure is now found to point in a diagonal direction. Further increasing
$r_c$ to large values $r_c/a\gg 1$ progressively reduces the stripe
anisotropy, in favor of a triangular cluster crystal, similar to the
continuous case~\cite{Cinti2010}.\\

\begin{figure}
    \centering
    \includegraphics{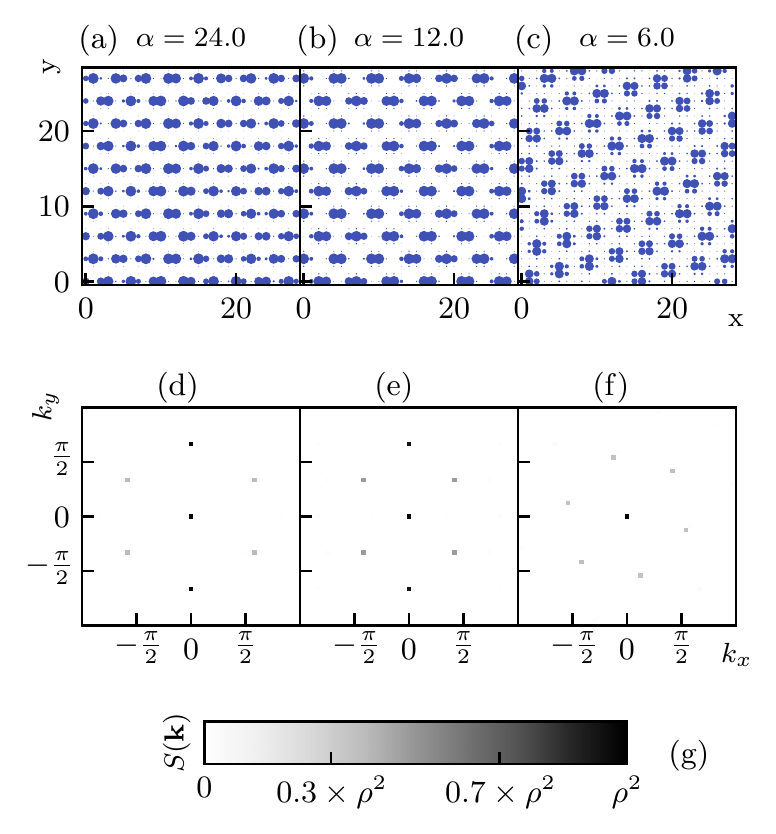}
    \caption{\label{fig:maps}%
        Panels (a)-(c): site-density maps of a \(48\times48\) system for
        different values of \(\alpha\). The size of each dot is propotional to
        the average occupation of the corresponding site.  Calculation are
        performed at \(T/t = 1/20\), $\rho=5/36$, and \(V/t = 10, 10, 12\)
        respectively. Panels (d)-(f): Heatmap plot for the structure factor
        \(S(\mathbf{k})\) for the same calculations as panels (a)-(c). The
        intensity-color scale is shown in panel (g).
    }
\end{figure}

Similar effects can be obtained by considering a smoother potential, instead
of the stepwise potential of \paperhamiltonian{}. As an example,
Fig.~\ref{fig:maps} shows results for a potential of the type
$V/[1+(r/r_c)^\alpha]$, for different values of $\alpha$ and $V/t\gg1$.  For
$\alpha \rightarrow \infty$ the shape of this interaction  approximates well
the one in \paperhamiltonian{}. We find that for $\alpha \gtrsim 10$
stripe order is favoured along the $x$-direction ($y$-direction) for $V/t\gg
1$, similar to the discussion for \paperhamiltonian{}. For
$\alpha\lesssim 10$, however, the stripes progressively give way to a more
isotropic, triangular crystalline structure.  In particular, for $\alpha=6$
(relevant for Rydberg dressed atoms) the density map [panel (c)] and structure
factor [panel (f)] indicate a triangular crystal structure that is essentially
isotropic in the $x$ and $y$ directions.  This can be qualitatively explained
by considering that lower values of $\alpha$ effectively include more sites to
the edges of the interaction volume for each particle, different from the case
of \paper{}.
\footnote{%
    We note that for the case of smooth potentials as in Rydberg-dressed gases
    anisotropy can be re-introduced in the interactions  by considering
    anisotropic Rydberg states (e.g., excited Rydberg $p$-type states instead
    of $s$ states, as in~\cite{Zeiher2016, Glaetzle2015, vanBijnen2015}).
}\\

We find that strong metastable effects are often obtained in the numerical
simulations for the situations discussed above in the intermediate regime of
interactions $V/t\gtrsim 1$. The phase diagram has to be investigated on a
case-by-case basis.

\bibliography{library.bib}